# Comparing name generator designs in rural panel studies: analyzing alter retention and change

**Marian-Gabriel Hâncean**[1*]
[1]Department of Sociology, University of Bucharest, Bucharest, 050663, Romania
E-mail: gabriel.hancean@sas.unibuc.ro

**Jürgen Lerner**[2]
[2]Department of Computer and Information Science, University of Konstanz, Konstanz, 78457, Germany
E-mail: juergen.lerner@uni-konstanz.de

**Christopher McCarty**[3]
[3]Department of Anthropology/Bureau of Economic and Business Research, University of Florida, Gainesville, 32611, USA
E-mail: ufchris@ufl.edu

**Abstract**

We conducted a two-wave personal network study in a rural Romanian community, interviewing the same participants (n = 68) using two name generators. Wave 1 employed a fixed-choice generator (n = 25) focused on emotional closeness; Wave 2 used a free-choice generator based on frequent interaction. We compared tie characteristics and assessed retention across waves. Alters who were kin, co-residents, or emotionally close were more likely to be retained, regardless of generator type. These findings underscore the role of relational attributes in personal network stability and highlight design considerations for network studies in resource-limited, culturally distinct settings.

[*]Corresponding author:
Full name: Marian-Gabriel Hâncean
Address: 90 Panduri, Bucharest, 050663, Romania
Email: gabriel.hancean@sas.unibuc.ro

**Introduction**

Personal networks or the social ties individuals maintain across diverse relational domains are essential for understanding access to support, information, and resources. These networks encompass enduring bonds and context-specific interactions across kinship, friendship, work, and community (Fischer, 1982; Wellman, 1999). Their measurement is particularly critical in settings where informal relationships filter access to care, employment, and social identity. A typical personal network design includes the ego (focal individual), alters (social contacts), and the ties among them (McCarty, 2002; Wellman, 2007). Ego-alter ties vary by strength, frequency, and intimacy, while alter-alter ties indicate network cohesion. Network data are classified into structural variables (e.g., density, cliques, centralization) (Maya-Jariego, 2021) and compositional variables describing ego and alter characteristics (e.g., smoking prevalence, mean age, tie intensity) (McCarty et al., 2007).

The questions designed to elicit subsets of an ego's wider social universe (name generators) are the primary tool for collecting personal network data (Bernard et al., 1990). Given that personal networks may include over 1,700 known individuals (Killworth et al., 1990), subsetting is necessary to reduce respondent burden (McCarty et al., 2007). Generators constrain elicitation through criteria such as content (e.g., important matters) (Marsden, 1987), relational role (e.g., support ties) (Wellman & Wortley, 1990), or geography (Hâncean et al., 2021), often combined with time frames (Lubbers et al., 2010) and numeric limits, e.g., five (Marsden, 1987), twenty-five (Hâncean, Lerner, Perc, et al., 2025), or forty-five alters (Lubbers et al., 2010). These constraints aim to enhance cooperation, decrease respondent burden, and improve data accuracy.

Name generators typically follow two formats: fixed and free-choice. Fixed-choice generators prompt respondents to list a set number of alters, often prioritizing close ties (Campbell & Lee, 1991; Maya-Jariego, 2018). Free-choice generators may rely on interaction criteria, such as contact frequency, without nomination caps (Bidart & Charbonneau, 2011). The choice between formats influences both the structural and compositional characteristics of the networks elicited (Neal & Neal, 2017; Perry et al., 2018).

It is noteworthy that name generators pose various methodological challenges. Recall bias tends to favor strong, salient ties and underrepresent peripheral alters (Bell et al., 2007; Brewer, 2000; Marin, 2004). Interpretive variability across social positions and cultures affects validity (Bearman & Parigi, 2004). Interviewer and question-order effects can further influence the number and type of alters reported (Eagle & Proeschold-Bell, 2015; Marsden, 2003; Pustejovsky & Spillane, 2009). Different generators yield distinct network profiles (Marin & Hampton, 2007), and alternatives such as position generators (Lin & Dumin, 1986), resource generators (Van Der Gaag & Snijders, 2005), and participant-aided sociograms (Hogan et al., 2007) have been proposed. Additionally, respondents exhibit varied behaviors during alter elicitation, including cooperation (naming alters as instructed), recall limitations (forgetting eligible contacts), cognitive burden (fatigue during listing), and non-cooperativeness (avoiding or resisting the name generator task) (González-Casado et al., 2025).

Most evidence on name generators derives from cross-sectional comparisons across distinct samples, leaving little understanding of how network data shift when the same individuals respond to different generators over time; a limitation that constrains insights into network stability and instrument sensitivity. To address this gap, we conducted a panel study in a rural Romanian community, applying two distinct name generators to the same participants at two-time points. In Wave 1, a fixed-choice generator asked respondents to list 25 alters, giving priority to emotionally close alters. In Wave 2, a free-choice, interaction-based generator prompted them to name daily and weekly interaction partners across domains such as household, work, and church. This longitudinal design allows us to compare networks elicited by each method, distinguish persistent from context-dependent ties, and assess how elicitation strategies influence network composition. Namely, we examine: (1) differences in alter composition and relational attributes across generators; (2) the temporal stability of nominations, including tie retention and turnover; and (3) the sociodemographic and relational profiles of retained, dropped, and new alters. Our findings contribute to understanding how elicitation strategies shape the observed structure and stability of personal networks in longitudinal, resource-constrained, and culturally specific contexts.

**Methods**

*Study setting and sampling*

We conducted this study in Lerești, a rural commune (administrative subdivision) located approximately 160 km northwest of Bucharest, Romania. Lerești shares structural features with many rural communities across Romania, including limited healthcare access, digital disparities, and elevated poverty risk, compounded by high rates of labor migration that fragment households and reshape social support. These characteristics make Lerești a strategic site for examining personal networks under demographic stress, economic precarity, and social reconfiguration. Our data and study are part of a larger research project investigating the association between personal networks and social risk factors for cancer (https://4p-can.eu/).

At the time of data collection, Lerești had 4,330 residents, including 3,808 adults (aged 18+). We administered face-to-face structured interviews at two-time points, September 2023 and March 2024, six months apart, to capture potential behavioral change while maintaining cohort stability. This design allowed us to track three alter classes: dropped alters (named only in Wave 1), retained alters (named in both waves), and new alters (named first in Wave 2).

We employed a panel design and selected the cohort using link-tracing sampling (Hâncean et al., 2021; Hâncean, Lerner, Perc, et al., 2025; Mihăilă et al., 2024; Oană et al., 2024), in which each participant was invited to nominate and recruit up to five others. Interviews were conducted only after referral willingness was confirmed. Six initial seeds were purposively selected to maximize variation: sex (four men, two women), age (M = 54.3 years, range = 30), income (three above and three below the national net average), education (five with university degrees, one with secondary education), and employment status (one retired, three employed in the public sector, one in the private sector, and one self-employed). We selected the seeds based on their availability and willingness to facilitate fieldwork; consequently, their sociodemographic characteristics do not necessarily reflect the distribution of population parameters.

The data collection and the study were conducted as part of the 4P-CAN project (HORIZON-MISS-2022-CANCER-01, Project ID: 101104432), a broader initiative focused on cancer prevention and health equity. In Wave 1, we interviewed 83 adults. To ensure diversity in sociodemographic characteristics, we stratified participants by age, sex, education, income, and employment status. In Wave 2, we successfully re-interviewed 68 participants from the original sample, yielding an attrition rate of 18.1%. Participants did not receive financial incentives but were offered access to a free medical hotline and invited to attend community-based cancer prevention events. Flexible scheduling and the relevance of the study context likely contributed to participant retention.

To assess recruitment dynamics within referral chains by age, sex, and education, we fitted exponential random graph models, i.e., ERGMs (Lusher et al., 2012). Full ERGM estimates and the socio-demographic profile of the study participants are available in the Supplementary Material (Hâncean, Lerner, & McCarty, 2025). **Table 1** presents the characteristics of the cohort egos together with available population-level estimates to support contextual interpretation.

All research procedures complied with the Declaration of Helsinki and the European General Data Protection Regulation (GDPR). The study protocol was approved by the Ethics Committee of the Center for Innovation in Medicine (EC-INOMED Decision No. D001/09-06-2023 and No. D001/19-01-2024). Written informed consent was obtained from all participants. We anonymized personal identifying information after each interview and securely stored data on encrypted drives accessible only to authorized personnel.

**Table 1. Sociodemographic characteristics of cohort egos (n = 68)**. This table presents the self-reported sociodemographic characteristics of study participants (egos) who completed interviews in both waves. Reported attributes include sex, age, education, occupational status, and personal income. Frequencies and percentages are provided for each category alongside available estimates from population-level data for contextual comparison.

| Variable, missing cases | Category | Cohort n (%) | Population N (%) |
|---|---|---|---|
| Sex, 0 | Male | 33 (49) | 1,790 (47) |
|  | Female | 35 (51) | 2,017 (53) |
| Age, 0 |  |  |  |
|  | 18-34 | 9 (13) | 724 (19) |
|  | 35-54 | 23 (34) | 1,333 (35) |
|  | 55-64 | 12 (18) | 609 (16) |
|  | 65+ | 24 (35) | 1,104 (29) |
| Education, 0 | Low (elementary) | 2 (3) | - |
|  | Medium (high school) | 35 (51) | - |
|  | High (faculty) | 31 (46) | - |
| Occupation, 0 | Employed | 36 (53) | - |
|  | Unemployed | 4 (6) | - |
|  | Retired | 28 (41) | - |
| Personal income, 1 | < Minimum net salary | 15 (22) | - |
|  | Between minimum and average salary | 45 (66) | - |
|  | > Average net salary | 7 (10) | - |

*Data collection*

Seven trained interviewers followed standardized protocols and conducted, in Romanian, face-to-face interviews using Network Canvas software (Version 6.5.3) (Birkett et al., 2021). We collected ego- and alter-level attributes, along with ego–alter and alter–alter tie data. Demographic attributes (age, sex, education) and relational indicators (kinship, co-residence, emotional closeness, interaction frequency) were recorded for alters. Ego-level data included demographics and self-reported measures on cancer risk factors (smoking, diet, physical activity, sunscreen use, and BMI estimation).

In Wave 1, we used a fixed-choice name generator that asked participants to list 25 individuals aged 18+ with whom they *regularly* interact, giving priority to people they feel emotionally close to. *Please provide the names of 25 people you regularly talk to or meet who are at least 18 years old. These can be family members, friends, acquaintances, neighbors, co-workers, etc. Please begin with the people you feel closest to emotionally, then continue with those you consider close, and so on.* Although structural analysis was not the primary focus, we fixed the name generator at 25 alters as prior research shows this threshold preserves key network properties while minimizing fatigue (McCarty et al., 2007).

In Wave 2, we used a free-choice, interaction-based name generator that asked participants to nominate individuals they saw *daily* or *weekly*. *Please nominate people in your life that you see daily or weekly. These can be family you are living with, other relatives, co-workers (in case you work), people with whom you go to barbecues, with whom you go to various community events (like church), who are your neighbors, whom you visit, who visit you, etc.*

Emotional closeness (3 - *very close*, 2 - *close*, 1 - *less close*, 0 - *not at all close*) and meeting frequency (7 - *daily*, 6 - *weekly*, 5 - *every two weeks*, 4 - *once a month*, 3 - *a few times a year*, 2 - *once a year*, 1 - *less than once a year*) were assessed only in Wave 1, consistent with the affective framing of the fixed-choice generator. We excluded these variables from the Wave 2 protocol to minimize burden and because the interaction-based generator prioritized frequency-based recall.

We included random effects in the statistical models to account for the nested data structure. We specified random intercepts for ego identity (and wave for pooled analyses). Interviewer effects were tested but excluded due to negligible variance.

While not central to the present analysis, we revised the name generator between Wave 1 and Wave 2 to improve the accuracy of ego-reported information on alters. This adjustment supported our broader objective within the 4P-CAN project: to examine how personal networks relate to cancer-related social risk factors by enhancing the report of socially and behaviorally relevant ties.

*Measurements*

To assess network change across waves, we computed a series of ego-level metrics based on alter nominations. We matched alters across waves using unique identifiers, enabling classification as retained, dropped, or new.

For each ego, we computed several key metrics. (1) *Stability rate*: proportion of Wave 1 alters retained in Wave 2. (2) *Turnover rate*: proportion of total alters that were either dropped or newly added relative to the union of alters across waves. High values suggest dynamic or unstable networks. (3) *Delta*: the numeric difference in network size (Wave 2 minus Wave 1), used to assess directional shifts in network volume. (4) *Jaccard index*: the ratio of shared alters to the total number of unique alters across both waves, quantifying overall network overlap. (5) *Gain/Loss ratio*: the ratio of new to dropped alters, offering insight into whether network change is additive, subtractive, or balanced. (6) *Proportion of new alters* (Wave 2): fraction of Wave 2 alters that were not listed in Wave 1. (7) *Proportion of dropped alters* (Wave 1): fraction of Wave 1 alters that did not reappear in Wave 2.

Although Wave 1 employed a fixed-choice name generator instructing participants to enumerate 25 alters, 13 of the 68 respondents did not reach this target. We kept these cases in the analysis to avoid unnecessary data loss and to preserve analytically valuable variation in reporting behavior.

*Model structure and random effects*

To examine predictors of alter retention, we estimated three mixed-effects logistic regression models with binary outcomes. The pooled model combined data from both waves and contrasted retained alters with those dropped or newly added (retained vs. dropped/new). Wave-specific models estimated retention separately for Wave 1 (retained vs. dropped) and Wave 2 (retained vs. new). All models included random intercepts for ego identity (n = 68) to account for within-ego clustering; the pooled model also included a random intercept for wave (n = 2). We initially included interviewer identity as an additional random effect, but its variance component was negligible, indicating minimal interviewer-related clustering. For parsimony, the final models retained random effects for ego (and wave, where applicable). Fixed effects included alter and ego demographics (age, sex, education) and tie-level attributes (family tie, co-residence, emotional closeness, and meeting frequency, where available). Full model diagnostics and replication materials are provided in the Supplementary Material. All analyses were conducted in RStudio (version 2024.12.1+563).

**Results**

*Descriptive statistics*

The fixed-choice generator (Wave 1), designed to give priority to emotionally salient ties, produced networks with a higher proportion of kin (44.0%) and emotionally close alters (30.6%, emotionally *close* and *very close*). It also yielded a majority of alters reported as being seen *daily* or *weekly* (54.9%), reflecting its emphasis on sustained emotional and interactional presence. In contrast, the non-fixed, interaction-based generator (Wave 2), which asked about daily/weekly contact in specific social domains rather than closeness, captured a more diverse set of ties, including a lower proportion of kin (25.9%) but a higher proportion of co-resident alters (10.7%). These patterns align with the focus of each elicitation strategy: the fixed-choice generator emphasized emotionally significant and socially close ties, while the interaction-based generator broadened the scope to include structurally proximal, but potentially less affectively bonded, network members.

Descriptive statistics for ego-level alter nominations are presented in **Table 2**. On average, egos nominated approximately 24 alters in Wave 1 (SD = 2.62), 17 in Wave 2 (SD = 6.38), and 32 unique alters named across both waves (SD = 5.58). Also, the number of alters named only in Wave 2 exhibited the highest relative variability (CV = 60.65%), indicating substantial variability across egos in the number of new alters introduced. This pattern contrasts with the more stable reporting in Wave 1 (CV = 10.88%) and the moderate variation in total unique alters (CV = 17.23%). Similarly, the SD-to-median ratio for Wave 2 nominations (0.38) and new alters (0.72) suggests that elicitation might have been more inconsistent in the second wave. These metrics collectively support that changing the name generator influenced egos' behavior, particularly by increasing variability in new nominations.

**Table 2. Summary of ego-level network change between waves**. Summary statistics describing alter nominations across survey waves (n = 68 egos). In addition to central tendency and dispersion measures, we report the coefficient of variation (CV) and the ratio of standard deviation to the median. These metrics highlight variability in reporting.

| Category | Mean | Median | SD | CV % | SD to Median |
|---|---|---|---|---|---|
| Alters in both waves (retained) | 9.03 | 8.5 | 4.37 | 48.37 | 0.51 |
| Only in Wave 1 (dropped) | 15.01 | 16.0 | 4.54 | 30.22 | 0.28 |
| Only in Wave 2 (new) | 8.35 | 7.0 | 5.07 | 60.65 | 0.72 |
| Total alters in Wave 1 | 24.04 | 25.0 | 2.62 | 10.88 | 0.10 |
| Total alters in Wave 2 | 17.38 | 17.0 | 6.38 | 36.71 | 0.38 |
| Total unique alters across waves | 32.40 | 31.5 | 5.58 | 17.23 | 0.18 |

The distributions of ego-level network stability and change measurements across waves (**Figure 1**) reveal variation between individuals in network stability over time (from Wave 1 to Wave 2). The mean *stability rate* is 0.38 (SD = 0.17), suggesting a limited overlap in alter composition. Consistent with this, the mean *Jaccard index* is 0.28 (SD = 0.14). These low values indicate that most egos reported substantially different sets of alters between waves. The average *network change* (*delta*) was −6.66 alters (SD = 6.60), indicating a consistent reduction in reported network size. On average, egos listed fewer alters in Wave 2. This contraction is further supported by the *gain/loss ratio* (M = 0.64, SD = 0.52), indicating that more alters were dropped than added between waves. The *turnover rate*, which reflects the total proportion of alters either lost or gained, averaged 0.72 (SD = 0.14). **This indicates that nearly three-quarters of reported alters were not consistently named across waves.**

Moreover, the proportion of *new alters introduced in Wave 2* was 0.46 (SD = 0.20), while *dropped alters from Wave 1* accounted for 0.62 (SD = 0.17), underscoring asymmetrical recall (i.e., an imbalance in the likelihood that participants remember and report the same alters across repeated survey waves) or cooperation. On average, 46% of alters reported in Wave 2 were new, not mentioned previously. Meanwhile, 62% of Wave 1 alters were dropped and not elicited in the second wave. This imbalance suggests that participants were more likely to omit previously mentioned alters than to name them consistently across waves. Alternatively, they may have interpreted the name generator differently in Wave 2, introducing asymmetry into the elicitation process. (We report the specific distributions of these measurements per ego in the Supplementary Material).

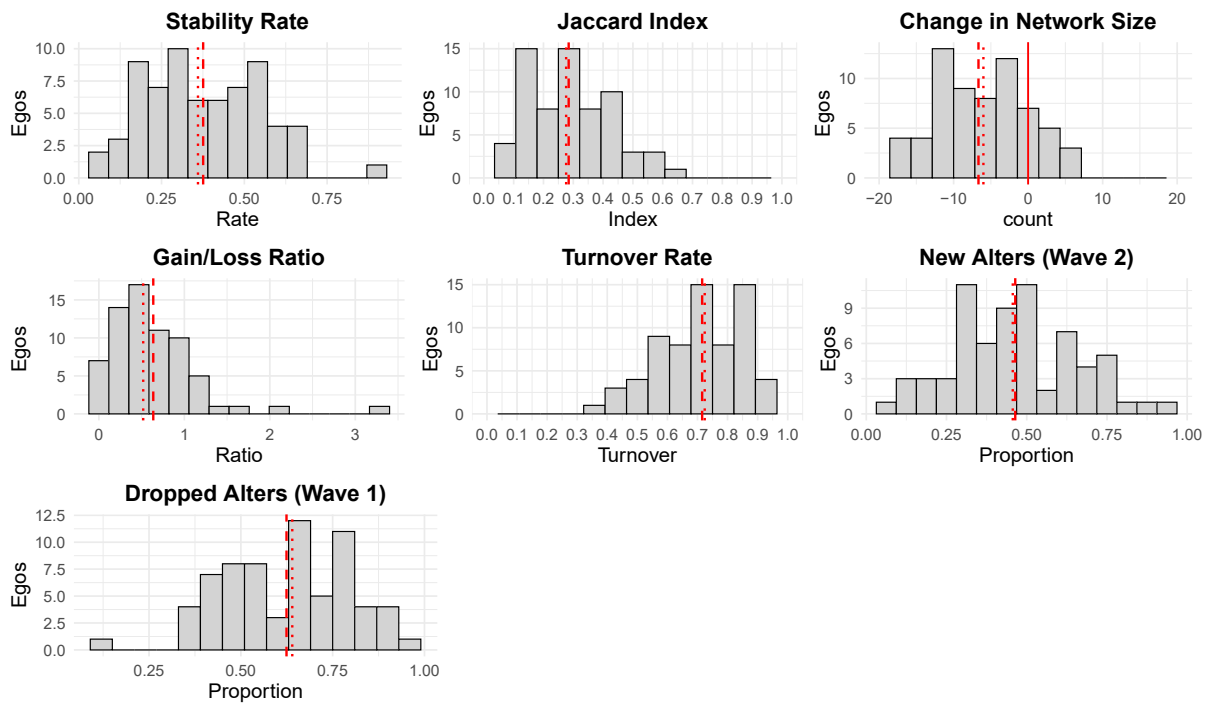

**Figure 1. Distributions of Ego-Level network stability and change measurements across waves.** Each histogram reflects a distinct dimension of change: (1) Stability rate (proportion of Wave 1 alters retained in Wave 2), (2) Jaccard index (overlap between waves), (3) Delta or net change in alter count, (4) Gain/Loss ratio (relative volume of additions vs. losses), (5) Turnover rate (total replacement activity), (6) number of new alters in Wave 2, and (7) Dropped alters from Wave 1. Vertical dashed and dotted lines indicate the mean and median for each distribution.

*Descriptive profile of alters based on retention status*

**Table 3** provides a descriptive overview of alters and ego-alter ties, disaggregated by sample status across waves: Dropped (named in Wave 1 but not re-elicited in Wave 2), Retained (named in both waves), and New (first elicited in Wave 2). Alters across these three groups were of similar age, with means ranging narrowly from 53.1 to 54.5 years and standard deviations around 16 years, indicating moderate age variability. Sex composition was consistent, with female representation ranging from 53.7% to 56.3% across categories and coefficients of variation (CVs) indicating relatively low dispersion. Educational distributions were comparable across groups, with medium-level education (high-school) being the most prevalent (52.8% - 58.7%). The proportion of alters with low educational attainment (lower secondary) was highest among newly reported alters (6.2%), and the high coefficient of variation (CV = 3.87) indicates considerable variation in educational backgrounds within this subgroup. High education (faculty) was evenly distributed across categories, comprising approximately 37.0% - 41.0% of alters.

Ego-alter tie characteristics demonstrated greater differentiation by status. Kinship ties were most prevalent among retained alters (53.7%) and less common among dropped (38.2%) and new alters (16.0%), suggesting stronger role-based salience in sustained network membership. Co-residence followed a similar pattern, occurring most frequently among retained alters (14.0%) while remaining infrequent among dropped (0.8%) and new (7.0%) alters. Measures of interaction quality (i.e., meeting frequency and emotional closeness) were only available for Wave 1, yet revealed meaningful contrasts. Retained alters were seen most frequently (M = 6.1, SD = 1.2) and rated as more emotionally close (M = 2.4, SD = 0.7) than dropped alters (M = 4.6, SD = 1.7 for meeting; M = 1.9, SD = 0.8 for closeness).

These results suggest that alters who remained in respondents' networks over time were not only more likely to be structurally close (e.g., family, co-residing) but also perceived as relationally closer. Conversely, dropped and new alters were more peripheral, both demographically and relationally, underscoring the selective retention of proximal and emotionally significant ties in the longitudinal egocentric network data. Further analyses indicate that alters named earlier in the interview tended to be emotionally closer to the ego, particularly when the alter was a family member (see Supplementary Material). This pattern suggests that closeness, especially kinship-based, influences elicitation order, with egos prioritizing socially salient ties when responding to name generators.

**Table 3. Demographic and relational profile of alters and ego-alter ties by sample status across waves**
Descriptive statistics are presented for alters and ego-alter relational characteristics based on sample status: Dropped (alters named in Wave 1 but not re-elicited in Wave 2), Retained (named in both waves), and New (first elicited in Wave 2). Age is reported as mean and standard deviation (SD). Proportions for sex, education (low, medium, high), family ties, and co-residence are shown as percentages with Coefficient of Variation (CV). Meeting frequency and emotional closeness were assessed only in Wave 1 and are reported as means and SDs.

| *Alters* | Dropped | Retained | New |
|---|---|---|---|
| n | 1021 | 614 | 568 |
| Age: M (SD) | 53.5 (15.9) | 53.1 (16) | 54.5 (16.8) |
| Female: % (CV) | 54.1% (0.92) | 53.7% (0.93) | 56.3% (0.88) |
| Education | | | |
|   Low: % (CV) | 3.5% (5.27) | 4.3% (4.72) | 6.2% (3.87) |
|   Medium: % (CV) | 55.9% (0.89) | 58.7% (0.84) | 52.8% (0.95) |
|   High: % (CV) | 40.6% (1.21) | 37.0% (1.3) | 41.0% (1.2) |
| *Ego-alter ties* | | | |
| Family tie: % (CV) | 38.2% (1.27) | 53.7% (0.93) | 16% (2.29) |
| Co-living: % (CV) | 0.8% (11.25) | 14% (2.48) | 7% (3.63) |
| Meeting freq.: M (SD) | 4.6 (1.7) | 6.1 (1.2) | NA |
| Emotional closeness: M (SD) | 1.9 (0.8) | 2.4 (0.7) | NA |

*Predicting retained alters*

We estimated three mixed-effects logistic regression models to examine predictors of alter retention across the two waves (**Table 4**). The pooled mixed-effects model (n = 2,738; AIC = 3360.1; retention vs. dropped/new) included random intercepts for both ego identity and study wave to account for clustering within networks and across time points. The model examined the effects of alter-level and ego-level covariates on the likelihood of alter retention. Family ties (OR ≈ 2.23, p < .001) and co-residence (OR ≈ 6.45, p < .001) emerged as strong and consistent predictors of alter retention. Alters who were family members or co-resided with the ego were substantially more likely to be retained across study waves. In contrast, older egos were significantly less likely to retain alters (p = .003), suggesting that age-related transitions may contribute to changes in personal network composition. No significant associations were observed for alter age, sex, education, or ego-level education and sex. The Wave 1 model (n = 1,583; AIC = 1591.8; retention vs. dropped) extended this specification by incorporating tie-level variables that were available only at baseline. Here, in addition to family and co-residence effects, contact frequency (OR ≈ 2.05, p < .001) and emotional closeness (OR ≈ 2.33, p < .001) were both positively associated with retention, reinforcing the salience of relational strength in network persistence. The Wave 2 model (n = 1,155; AIC = 1451.1; retention vs. new) reaffirmed the predictive power of family ties (OR ≈ 4.13, p < .001) and co-residence (OR ≈ 4.87, p < .001). Ego age again exhibited a negative association with alter retention (p = .013). Further analyses (see Supplementary Material) show that older egos do not significantly reduce family ties over time but are more likely to drop non-family alters between waves.

**Table 4. Comparative summary of retention predictors across logistic models**

This table reports estimated coefficients, standard errors, odds ratios, and p-values from three mixed-effects logistic regression models predicting alter retention in longitudinal personal networks. The pooled model includes observations from both waves and contrasts retained alters with those dropped or newly added. Wave-specific models compare retained alters with dropped alters (Wave 1) and newly added alters (Wave 2). Predictor variables include alter and ego demographics (age, sex, education) and tie-level attributes (family tie, co-residence, closeness, and meeting frequency, where available). Statistically significant estimates are denoted by asterisks (p < .05, p < .01, p < .001). All models include random intercepts for ego (n = 68); the pooled model also includes a random intercept for wave (n = 2). Model fit indices (AIC, BIC, pseudo-$R^2$) and number of observations are reported for each specification. Variance components for random effects are shown in a panel below the fixed-effect estimates.

| Predictor | Pooled Est. (SE) OR p-value | Wave 1 Est. (SE) OR p-value | Wave 2 Est. (SE) OR p-value |
|---|---|---|---|
| | *Retained vs. Dropped/New* | *Retained vs. Dropped* | *Retained vs. New* |
| (Intercept) | 0.061 (0.736) 1.063 p = .934 | -7.047 (0.967) 0.001 p < .001 *** | 0.771 (0.861) 2.162 p = .371 |
| **Alter attributes** | | | |
| Age | 0.004 (0.003) 1.004 p = .272 | -0.000 (0.005) 1.000 p = .922 | 0.002 (0.005) 1.002 p = .736 |
| Education | -0.007 (0.084) 0.993 p = .930 | -0.026 (0.131) 0.974 p = .845 | 0.053 (0.128) 1.054 p = .677 |
| Sex | -0.028 (0.091) 0.972 p = .755 | -0.048 (0.139) 0.953 p = .729 | -0.064 (0.139) 0.938 p = .642 |
| **Ego attributes** | | | |
| Age | -0.019 (0.006) 0.981 p = .003 ** | -0.006 (0.008) 0.994 p = .436 | -0.020 (0.008) 0.980 p = .013 * |
| Education | 0.116 (0.167) 1.123 p = .487 | 0.348 (0.201) 1.416 p = .083 | -0.034 (0.203) 0.967 p = .867 |
| Sex | -0.076 (0.185) 0.927 p = .679 | 0.168 (0.224) 1.183 p = .452 | -0.049 (0.224) 0.952 p = .828 |
| **Ego-alter tie** | | | |
| Colive | 1.863 (0.203) 6.443 p < .001 *** | 1.381 (0.420) 3.979 p = .001 ** | 1.583 (0.252) 4.870 p < .001 *** |
| Family | 0.803 (0.094) 2.232 p < .001 *** | 0.323 (0.154) 1.381 p = .037 * | 1.419 (0.164) 4.133 p < .001 *** |
| Closeness | – | 0.844 (0.118) 2.326 p < .001 *** | – |
| Meeting freq. | – | 0.719 (0.054) 2.052 | – |

|                        |                 | *p* < .001 ***  |                 |
|------------------------|-----------------|-----------------|-----------------|
| **Random effects**     | *Variance (SD)* | *Variance (SD)* | *Variance (SD)* |
| Ego ID (n = 68)        | 0.420 (0.648)   | 0.494 (0.703)   | 0.479 (0.692)   |
| Wave (n=2)             | 0.123 (0.350)   | –               | –               |
|                        |                 |                 |                 |
| **Model fit**          |                 |                 |                 |
| AIC                    | 3,360.1         | 1,591.83        | 1,451.12        |
| BIC                    | 3,425.2         | 1,656.24        | 1,501.64        |
| R² marginal            | 0.13            | 0.43            | 0.13            |
| R² conditional         | 0.23            | 0.50            | 0.24            |
| Observations           | 2738            | 1583            | 1155            |

**Discussion**

Our panel study examined how two different name generator strategies, i.e., one fixed-choice giving priority to emotional closeness and one free-choice emphasizing frequent interaction, shape the elicitation and longitudinal retention of alters in personal network interviews. By administering both generators to the same participants across two waves, we directly compared their elicited networks within individuals in a rural Romanian setting. Consistent with prior findings (Bidart & Charbonneau, 2011), the fixed-choice generator yielded networks centered around emotionally close ties, whereas the interaction-based generator captured a broader, more heterogeneous set of alters.

However, our findings demonstrate that the stability of personal networks over time was more closely associated with relational characteristics, i.e., kinship and co-residence, than with the elicitation strategy per se. This result aligns with prior work (Koster, 2018), that claimed the enduring role of kin-based ties in rural contexts. Despite the broader reach of the free-choice generator, alters retained across waves were disproportionately embedded in strong, proximal roles, echoing insights from studies on the structural stability of rural networks (Bignami-Van Assche, 2005).

Several limitations should be acknowledged. First, emotional closeness and meeting frequency were only measured in Wave 1 due to their embedding in the fixed-choice generator. This asymmetry constrains direct comparisons of tie strength across waves. Second, unobserved biases, such as varying interpretations of prompts, differential recall or cooperation may have influenced the networks elicited. Additionally, switching generator types likely introduced measurement errors due to differences in cognitive load and task interpretation. These factors may have affected consistency in alter reporting across waves and should be acknowledged when evaluating network stability.

Despite these constraints, the consistency of kinship and co-residence as predictors of tie retention underlines the importance of relational embeddedness in the elicitation process. These findings support previous work indicating that strong ties, such as those based on kinship or co-residence, anchor longitudinal network data and are more resistant to decay, cognitive load, or recall bias than weaker, more peripheral ties (Ureña-Carrion et al., 2020). In our study, the use of distinct name generators likely produced complementary views of personal networks. However, alter retention across waves was primarily driven by relational embeddedness.

Although context-specific, our approach offers methodological insights for longitudinal network research in rural or structurally constrained environments. Comparing name generator formats within a single cohort provides a rare opportunity to assess how instrument design interacts with social structure. We echo prior calls (Bidart & Charbonneau, 2011) for more integrated or tailored generator strategies and suggest that attention to tie attributes (kinship and proximity) is essential for promoting longitudinal stability and measurement comparability.

**Data availability**

The dataset analyzed in the current study and the R code (as Supplementary Material) are made openly available in the Zenodo data repository as Hâncean, M.-G., Lerner, J., & McCarty, C. (2025). Replication data for: Comparing name generator designs in rural panel studies. [Dataset]. Zenodo. https://doi.org/10.5281/zenodo.15322440.

**Acknowledgments**

M.-G.H. was supported by the 4P-CAN project, HORIZON-MISS-2022-CANCER-01, project ID 101104432, programme HORIZON; J.L. was supported by Deutsche Forschungsgemeinschaft (DFG 321869138). The funding sources had no involvement in study design, in the collection, analysis, and interpretation of data, in writing the paper, and in the decision to submit the article for publication. We express our gratitude to Marius Geantă, Cosmina Cioroboiu, Iulian Oană, Bianca-Elena Mihăilă, Florin Găină, Simona-Elena Puncioiu, Bogdan-Adrian Vidrașcu, Maria-Alexandra Roșu, Isabela Tincă and Maria-Cristina Ghiță for their contribution to this study.

# References

Bearman, P., & Parigi, P. (2004). Cloning headless frogs and other important matters: Conversation topics and network structure. *Social Forces*, *83*(2), 535–557. https://doi.org/10.1353/sof.2005.0001

Bell, D. C., Belli-McQueen, B., & Haider, A. (2007). Partner naming and forgetting: Recall of network members. *Social Networks*, *29*(2), 279–299. https://doi.org/10.1016/j.socnet.2006.12.004

Bernard, H. R., Johnsen, E. C., Killworth, P. D., McCarty, C., Shelley, G. A., & Robinson, S. (1990). Comparing four different methods for measuring personal social networks. *Social Networks*, *12*(3), 179–215. https://doi.org/10.1016/0378-8733(90)90005-T

Bidart, C., & Charbonneau, J. (2011). How to generate personal networks: Issues and tools for a sociological perspective. *Field Methods*, *23*(3), 266–286. https://doi.org/10.1177/1525822X11408513

Bignami-Van Assche, S. (2005). Network stability in longitudinal data: A case study from rural Malawi. *Social Networks*, *27*(3), 231–247. https://doi.org/10.1016/j.socnet.2005.02.001

Birkett, M., Janulis, P., Phillips, G. second, Contractor, N., & Hogan, B. (2021). Network Canvas: Key decisions in the design of an interviewer-assisted network data collection software suite. *Social Networks*, *66*, 114–124. https://doi.org/10.1016/j.socnet.2021.02.003

Brewer, D. D. (2000). Forgetting in the recall-based elicitation of personal and social networks. *Social Networks*, *22*(1), 29–43. https://doi.org/10.1016/S0378-8733(99)00017-9

Campbell, K. E., & Lee, B. A. (1991). Name generators in surveys of personal networks. *Social Networks*, *13*(3), 203–221. https://doi.org/10.1016/0378-8733(91)90006-F

Eagle, D. E., & Proeschold-Bell, R. J. (2015). Methodological considerations in the use of name generators and interpreters. *Social Networks*, *40*, 75–83. https://doi.org/10.1016/j.socnet.2014.07.005

Fischer, C. S. (1982). *To dwell among friends: Personal networks in town and city*. University of Chicago Press.

González-Casado, M. A., Rey, A. C., Corrotea, M. P., McCarty, C., Molina, J. L., & Sánchez, A. (2025). *Name generators in egocentric network research: A comparative analysis of three approaches*. [Preprint]. SocArXiv. https://doi.org/10.31235/osf.io/kumcj_v1

Hâncean, M.-G., Lerner, J., & McCarty, C. (2025). *Replication data for: Comparing name generator designs in rural panel studies*. [Dataset]. Zenodo. https://doi.org/10.5281/zenodo.15322440

Hâncean, M.-G., Lerner, J., Perc, M., Molina, J. L., Geantă, M., Oană, I., & Mihăilă, B.-E. (2025). Processed food intake assortativity in the personal networks of older adults. *Scientific Reports*, *15,* 10459. https://doi.org/10.1038/S41598-025-94969-0

Hâncean, M.-G., Lubbers, M. J., & Molina, J. L. (2021). Measuring transnational social fields through binational link-tracing sampling. *PLOS ONE*, *16*(6), e0253042. https://doi.org/10.1371/journal.pone.0253042

Hogan, B., Carrasco, J. A., & Wellman, B. (2007). Visualizing personal networks: Working with participant-aided sociograms. *Field Methods*, *19*(2), 116–144. https://doi.org/10.1177/1525822X06298589

Killworth, P. D., Johnsen, E. C., Bernard, H. R., Ann Shelley, G., & McCarty, C. (1990). Estimating the size of personal networks. *Social Networks*, *12*(4), 289–312. https://doi.org/10.1016/0378-8733(90)90012-X

Koster, J. (2018). Family ties: The multilevel effects of households and kinship on the networks of individuals. *Royal Society Open Science*, *5*(4), 172159. https://doi.org/10.1098/rsos.172159

Lin, N., & Dumin, M. (1986). Access to occupations through social ties. *Social Networks*, *8*(4), 365–385. https://doi.org/10.1016/0378-8733(86)90003-1

Lubbers, M. J., Molina, J. L., Lerner, J., Brandes, U., Ávila, J., & McCarty, C. (2010). Longitudinal analysis of personal networks. The case of Argentinean migrants in Spain. *Social Networks*, *32*(1), 91–104. https://doi.org/10.1016/j.socnet.2009.05.001

Lusher, D., Koskinen, J., & Robins, G. (Eds.). (2012). *Exponential random graph models for social networks: Theory, methods, and applications* (1st ed.). Cambridge University Press. https://doi.org/10.1017/CBO9780511894701

Marin, A. (2004). Are respondents more likely to list alters with certain characteristics? *Social Networks*, *26*(4), 289–307. https://doi.org/10.1016/j.socnet.2004.06.001

Marin, A., & Hampton, K. N. (2007). Simplifying the personal network name generator: Alternatives to traditional multiple and single name generators. *Field Methods*, *19*(2), 163–193. https://doi.org/10.1177/1525822X06298588

Marsden, P. V. (1987). Core discussion networks of Americans. *American Sociological Review*, *52*(1), 122. https://doi.org/10.2307/2095397

Marsden, P. V. (2003). Interviewer effects in measuring network size using a single name generator. *Social Networks*, *25*(1), 1–16. https://doi.org/10.1016/S0378-8733(02)00009-6

Maya Jariego, I. (2018). Why name generators with a fixed number of alters may be a pragmatic option for personal network analysis. *American Journal of Community Psychology*, *62*(1–2), 233–238. https://doi.org/10.1002/ajcp.12271

Maya-Jariego, I. (2021). Building a structural typology of personal networks: Individual differences in the cohesion of interpersonal environment. *Social Networks*, *64*, 173–180. https://doi.org/10.1016/j.socnet.2020.09.006

McCarty, C. (2002). Structure in personal networks. *Journal of Social Structure*, *3*(1), 1–36.

McCarty, C., Killworth, P. D., & Rennell, J. (2007). Impact of methods for reducing respondent burden on personal network structural measures. *Social Networks*, *29*(2), 300–315. https://doi.org/10.1016/j.socnet.2006.12.005

Mihăilă, B.-E., Hâncean, M.-G., Perc, M., Lerner, J., Oană, I., Geantă, M., Molina, J. L., & Cioroboiu, C. (2024). Cross-sectional personal network analysis of adult smoking in rural areas. *Royal Society Open Science*, *11*(11), 241459. https://doi.org/10.1098/rsos.241459

Neal, Z. P., & Neal, J. W. (2017). Network analysis in community psychology: Looking back, looking forward. *American Journal of Community Psychology*, *60*(1–2), 279–295. https://doi.org/10.1002/ajcp.12158

Oană, I., Hâncean, M.-G., Perc, M., Lerner, J., Mihăilă, B.-E., Geantă, M., Molina, J. L., Tincă, I., & Espina, C. (2024). Online media use and COVID-19 vaccination in real-world personal networks: Quantitative study. *Journal of Medical Internet Research*, *26*, e58257. https://doi.org/10.2196/58257

Perry, B. L., Pescosolido, B. A., & Borgatti, S. P. (2018). *Egocentric network analysis: Foundations, methods, and models* (1st ed.). Cambridge University Press. https://doi.org/10.1017/9781316443255

Pustejovsky, J. E., & Spillane, J. P. (2009). Question-order effects in social network name generators. *Social Networks*, *31*(4), 221–229. https://doi.org/10.1016/j.socnet.2009.06.001

Ureña-Carrion, J., Saramäki, J., & Kivelä, M. (2020). Estimating tie strength in social networks using temporal communication data. *EPJ Data Science*, *9*(1), 37. https://doi.org/10.1140/epjds/s13688-020-00256-5

Van Der Gaag, M., & Snijders, T. A. B. (2005). The resource generator: Social capital quantification with concrete items. *Social Networks*, *27*(1), 1–29. https://doi.org/10.1016/j.socnet.2004.10.001

Wellman, B. (Ed.). (1999). *Networks in the global village: Life in contemporary communities*. Westview Press.

Wellman, B. (2007). Challenges in collecting personal network data: The nature of personal network analysis. *Field Methods*, *19*(2), 111–115. https://doi.org/10.1177/1525822X06299133

Wellman, B., & Wortley, S. (1990). Different strokes from different folks: Community ties and social support. *American Journal of Sociology*, *96*(3), 558–588. https://doi.org/10.1086/229572